\documentstyle[aps,epsfig]{revtex}

\makeatletter

\begin{document}

{\par\centering {\huge Theory of angular correlation for double photoionization in a rare gas atom
due to linearly polarized light}\huge \par}
\vspace{0.5in}

{\par\centering {\large Chiranjib Sur and Dipankar Chattarji}\large \par}

{\par\centering \emph {Department of Physics, Visva-Bharati, Santiniketan 731 235, INDIA}\par}

\vspace{0.3in}
{\small In this paper we have tried to establish a theory for angular correlation between
the photo- and Auger electrons in the two-step double photoionization of a rare
gas atom. The basic finding of this paper is that a general theoretical expression
for the angular correlation function is established in the problem of double
photoionization of atoms in the presence of linearly polarized monochromatic
light. We consider the case of a xenon atom and the incident light is of \( 94.5\, eV \)
with arbitrary linear polarization. The theoretical results are compared with
the experiments done by K\( \ddot{a} \)mmerling and Schmidt {[}J.Phys.B, \textbf{26},
1141(1993){]}.{\small \par}
\vspace{0.2in}

\vspace{0.2in}

In a recent paper we gave a theory for angular correlation in the double photoionization
(DPI) of a rare gas atom {[}\ref{we1}{]}. We took DPI to be a two step process
{[}\ref{we2}{]}. The atom is irradiated by a beam of unpolarized photons and
emits a photo-electron. This is the first step. After a finite time interval
the singly ionized atom de-excites by emitting an Auger electron. The time delay
between the two steps decouples them energetically. However, the angular momenta
of the photo-electron and the Auger electron interact to yield their final angular
correlation. We defined the angular correlation \( W \) as a function of \( \theta  \),
the angle between the directions of emission of the two electrons.

The object of the present paper is to obtain a general expression for the angular
correlation function for DPI of a rare gas atom due to a linearly polarized
photon beam described by Stokes parameters \( S_{1} \), \( S_{2} \) and \( S_{3} \)
{[}\ref{born}{]}. The atom is initially randomly oriented, but becomes aligned
after the absorption of a photon. It is also energetically excited, and emits
a photo-electron from one of its inner shells. The photo-electron leaves the
interaction region in \( \sim 10^{-18} \) sec. The singly ionized atom with
an inner shell vacancy will now tend to de-excite by emitting an Auger electron
from one of the outer shells {[}\ref{dc}{]}. A typical Auger lifetime of say
the xenon (\( Z=54 \)) \( O \)-shell is \( \sim 10^{-15} \) sec. The photo-
and Auger ionizations are thus widely separated in time. 

We now come to the basic problem of calculating the angular correlation between
the photo-electron and the Auger electron. In a natural way, we can define the
angular correlation function \( W(\theta ) \) to be the probability that the
angle between their directions of emission is \( \theta  \). On what factors
should it depend, and how can we formulate a general approach to its investigation?
The following points may be noted.\\
(i) Obviously, \( W(\theta ) \) should be independent of the choice of coordinates,
because basically it is a function of the angle \( \theta  \). This calls for
a tensor formalism.\\
(ii) Since \( W \) has the nature of a probability, we can only evaluate it
by counting the number of coincident events in which the angle between the directions
of emission has a particular value \( \theta  \), and so on. Thus the question
arises of a set of detectors which can concurrently detect the photo-electron
and the Auger electron. In other words, we have to define a detection operator
to describe such operations quantum mechanically.\\
(iii) The detectors may or may not be completely dependable in actual practice.
This raises the question of efficiency of a particular detector.\\
(iv) Thus we can think of two kinds of quantum mechanical operators to describe
our problem: (a) a density operator \( \rho  \) which describes the probability
of occurrence of an electron in a particular state, and (b) an efficiency operator
\( \varepsilon  \) which describes the probability that our detector will actually
detect the electron in that state.

We can thus visualize the angular correlation function as a scaler which is
simply related to the product of the two tensor operators \( \rho  \) and \( \varepsilon  \)
as defined above. Evidently a contraction of these tensor operators will be
called for to give the scaler \( W(\theta ) \). One virtue of this theory is
that we are invoking the statistical properties of a complete ensemble and no
perturbation series of any kind is involved.

Let us now look a little more carefully at the angular correlation function
\( W(\theta ) \). When we say that it is the probability that the angle between
the directions of emission of the two electrons is \( \theta  \), we actually
mean two things. (i) The two electrons are in states which are compatible with
their directions of emission making an angle \( \theta  \) with each other,
and (ii) the efficiencies of the two electron detectors are such that the angle
between the directions of emission may be inferred to be \( \theta  \). Taking
these two factors into account we define

\begin{equation}
\label{1}
\overline{\varepsilon }=\sum _{J_{a}J^{\prime }_{a}\alpha _{a}\alpha ^{\prime }_{a}k_{a}\kappa _{a}}\rho _{k_{a}\kappa _{a}}(J_{a}\alpha _{a},J^{\prime }_{a}\alpha ^{\prime }_{a})\varepsilon ^{\star }_{k_{a}\kappa _{a}}(J_{a}\alpha _{a},J^{\prime }_{a}\alpha ^{\prime }_{a})\, ,
\end{equation}
 \( W(\theta ) \) being the angular part of \( \overline{\varepsilon } \).

As before {[}\ref{we1}{]}, we denote the initial state (photon+atom) by the
set of quantum numbers \( (J_{a}M_{a}\alpha _{a}) \), or by virtual quantum
numbers \( (J^{\prime }_{a}M^{\prime }_{a}\alpha ^{\prime }_{a}) \). \( (J_{a},M_{a}) \)
or \( (J^{\prime }_{a}M^{\prime }_{a}) \) are angular momentum quantum numbers,
and \( \alpha _{a} \),\( \alpha ^{\prime }_{a} \) stand for the set of remaining
quantum numbers. Similarly for the intermediate and final states. 

In DPI experiments, a Cylindrical Mirror Analyser (CMA) is used for detection
and energy analysis of the electrons. It imparts to the electron a cylindrical
symmetry regardless of its angle of incidence with respect to the axis. 

As for the orbital angular momentum of the incident photo- or Auger electron,
this is fixed by arranging to receive only those incident electrons which have
the characteristic energy. This implies that the detector is receiving all the
photo- or Auger electrons, as the case may be, having the characteristic energy
and orbital angular momentum. In other words, the CMA is acting not only as
a differential energy analyser, but also as an angle-integrated device. For
a given orbital angular momentum which labels a photo- or Auger electron, the
conjugate angle of incidence becomes completely indeterminate. It is this indeterminacy
which gives a statistical character to the distribution of the two electrons.
That in its turn manifests itself in their angular correlation.

We decompose the density matrix {[}\ref{blum}{]} of the photo-absorbed rare
gas atom in terms of the density matrix of the unobserved doubly ionized state
and of the two continuum electrons. The choice of detectors is discussed in
detail in reference {[}\ref{we3}{]}. The expression for the efficiency tensor
component is

\begin{equation}
\label{2}
\varepsilon _{k_{i}\kappa _{i}}^{\star }(j_{i}j^{\prime }_{i})=\sum _{\kappa ^{\prime }_{i}}z_{k_{i}}(i)c_{k_{i}\kappa ^{\prime }_{i}}(j_{i}j^{\prime }_{i})D_{\kappa ^{\prime }_{i}\kappa _{i}}^{k_{i}}(\Re _{i})\, .
\end{equation}
Since the residual doubly ionized state is unobserved we can write

\begin{equation}
\label{3}
\varepsilon _{k_{c}\kappa _{c}}^{\star }(J_{c}J^{\prime }_{c})=\sqrt{2J_{c}+1}\delta _{k_{c}0}\delta _{\kappa _{c}0}\delta _{J_{c}J^{\prime }_{c}}\, .
\end{equation}
 Some simplification yields the result

\begin{equation}
\label{4}
\begin{array}{cc}
\overline{\varepsilon } & =\sum \rho _{k_{a}\kappa _{a}}(J_{a},J^{\prime }_{a})\sqrt{2J_{c}+1}C_{\kappa _{b}\kappa _{1}\kappa _{a}}^{k_{b}k_{1}k_{a}}C_{0\kappa _{2}\kappa _{b}}^{0k_{2}k_{b}}\sqrt{2J_{a}+1}\sqrt{2J^{\prime }_{a}+1}\\
 & \times \sqrt{2k_{b}+1}\sqrt{2k_{1}+1}\sqrt{2J_{b}+1}\sqrt{2J^{\prime }_{b}+1}\sqrt{2k_{c}+1}\sqrt{2k_{2}+1}\\
 & \times \left\{ \begin{array}{ccc}
J_{c} & j_{2} & J_{b}\\
J^{\prime }_{c} & j^{\prime }_{2} & J^{\prime }_{b}\\
0 & k_{2} & k_{b}
\end{array}\right\} \left\{ \begin{array}{ccc}
J_{b} & j_{1} & J_{a}\\
J^{\prime }_{b} & j^{\prime }_{1} & J^{\prime }_{a}\\
k_{b} & k_{1} & k_{a}
\end{array}\right\} \\
 & \times z_{k_{1}}(1)c_{k_{1}\kappa _{1}}(j_{1}j^{\prime }_{1})z_{k_{2}}(2)c_{k_{2}\kappa _{2}}(j_{2}j^{\prime }_{2})D_{\kappa ^{\prime }_{1}\kappa _{1}}^{k_{1}}(\Re _{1})D_{\kappa ^{\prime }_{2}\kappa _{2}}^{k_{2}}(\Re _{2})\, .
\end{array}
\end{equation}

In experiments for measuring the angular correlation one usually chooses detectors
insensitive to the spin polarization of electrons. Then Eq.(\ref{4}) turns
out to be

\begin{equation}
\label{5}
\begin{array}{cc}
\overline{\varepsilon (\theta )} & \sim \sum (-1)^{J_{a}}\zeta\xi\left\{ \begin{array}{ccc}
J_{c} & j_{2} & J_{b}\\
J^{\prime }_{c} & j^{\prime }_{2} & J^{\prime }_{b}\\
0 & k_{2} & k_{b}
\end{array}\right\} \left\{ \begin{array}{ccc}
J_{b} & j_{1} & J_{a}\\
J^{\prime }_{b} & j^{\prime }_{1} & J^{\prime }_{a}\\
k_{b} & k_{1} & k_{a}
\end{array}\right\} z_{k_{1}}(1)z_{k_{2}}(2)\rho ^{\gamma }_{k_{a}\kappa _{a}}(1,1)\\
 & \times C_{000}^{k_{b}k_{1}k_{a}}C_{000}^{0k_{2}k_{b}}c_{k_{1}0}(j_{1}j^{\prime }_{1})c_{k_{2}0}(j_{2}j^{\prime }_{2})P_{k}(\cos \theta )\, ,
\end{array}
\end{equation}
 where \( \rho ^{\gamma }_{k_{a}\kappa _{a}}(1,1) \) is the density matrix
describing the photon, and 

\begin{equation}
\label{6}
\zeta=\sqrt{2J_{c}+1}\sqrt{2k_{a}+1}\sqrt{2k_{1}+1}\sqrt{2J_{b}+1}\sqrt{2J^{\prime }_{b}+1}\sqrt{2k_{2}+1}
\end{equation}
 and

\begin{equation}
\label{7}
\xi=\left\langle J_{b}\right\Vert j_{1}\left\Vert J_{a}\right\rangle \left\langle J_{b}\right\Vert j^{\prime }_{1}\left\Vert J_{a}\right\rangle ^{\star }\left\langle J_{c}\right\Vert j_{2}\left\Vert J_{b}\right\rangle \left\langle J_{c}\right\Vert j^{\prime }_{2}\left\Vert J_{b}\right\rangle ^{\star }.
\end{equation}

Now consider a randomly oriented \( ^{1}S_{0} \) xenon atom which is irradiated
by a \( 94.5\, eV \) photon beam of arbitrary linear polarization. This leads
to photoionization in the \( 4d_{5/2} \) shell followed by a subsequent \( N_{5}-O_{23}O_{23}\, ^{1}S_{0} \)
Auger decay {[}\ref{schmidt}{]}. We use the dipole approximation, and the letters
\( e,f \) and \( g \) for the three possible photoionization channels: \( e)4d_{5/2}\longrightarrow \varepsilon _{p}f_{7/2} \),
\( f)4d_{5/2}\longrightarrow \varepsilon _{p}f_{5/2} \) and \( g)4d_{5/2}\longrightarrow \varepsilon _{p}p_{3/2} \)
respectively. The Auger transition is characterised by the partial wave \( \varepsilon _{A}d_{5/2} \).
The same selection rules for photoionization and Auger transitions hold good
as in reference{[}\ref{we1}{]}. Since the photoionization process is described
by three possible channels, Eq.(\ref{5}) will be modified by the addition of
interference terms. The total intensity will, however, remain unchanged, as
we have discussed in reference {[}\ref{we1}{]}. Then the expectation value
of the efficiency operator turns out to be

\begin{equation}
\label{8}
\overline{\varepsilon (\theta )}=\overline{\varepsilon _{e}(\theta )}+\overline{\varepsilon _{f}(\theta )}+\overline{\varepsilon _{g}(\theta )}+\overline{\varepsilon _{ef}(\theta )}+\overline{\varepsilon _{fg}(\theta )}+\overline{\varepsilon _{ge}(\theta )}\, .
\end{equation}
 Here

\begin{equation}
\label{9}
\begin{array}{cc}
\overline{\varepsilon _{e}(\theta )} & \sim \sum z_{k_{1}}(1)z_{k_{2}}(2)\rho _{k_{a}\kappa _{a}}^{\gamma }(1,1)\left\{ \begin{array}{ccc}
J_{c} & j_{2} & J_{b}\\
J_{c} & j_{2} & J_{b}\\
0 & k_{2} & k_{b}
\end{array}\right\} \\
 & \times \left\{ \begin{array}{ccc}
J_{b} & j^{e}_{1} & J_{a}\\
J_{b} & j^{e}_{1} & J_{a}\\
k_{b} & k_{1} & k_{a}
\end{array}\right\} C_{000}^{k_{b}k_{1}k_{a}}C_{000}^{0k_{2}k_{b}}\\
 & \times \left| \left\langle J_{b}\right\Vert j^{e}_{1}\left\Vert J_{a}\right\rangle \right| ^{2}\left| \left\langle J_{c}\right\Vert j_{2}\left\Vert J_{b}\right\rangle \right| ^{2}c_{k_{1}0}(j^{e}_{1}j^{e}_{1})\\
 & \times c_{k_{2}0}(j_{2}j_{2})P_{k}(\cos \theta )\, ,
\end{array}
\end{equation}
 and the summation extends over \( k_{a},k_{1},k_{2} \) and \( k \). The expectation
values \( \overline{\varepsilon _{f}(\theta )} \) and \( \overline{\varepsilon _{g}(\theta )} \)
have the same form with \( \left\{ j^{e}_{i}\longrightarrow j^{f}_{1},\, k^{e}_{1}\longrightarrow k^{f}_{1}\right\}  \)
and \( \left\{ j^{e}_{i}\longrightarrow j^{f}_{1},\, k^{g}_{1}\longrightarrow k^{g}_{1}\right\}  \)
respectively. The quantity \( \overline{\varepsilon _{ij}(\theta )} \) is an
interference term arising from interaction between different photoionization
channels \( i \) and \( j \) (\( i,j=e,f,g \) with \( i\neq j \)).

\begin{equation}
\label{10}
\begin{array}{cc}
\overline{\varepsilon _{ij}(\theta )} & \sim \sum z_{k_{1}}(1)z_{k_{2}}(2)\rho _{k_{a}\kappa _{a}}^{\gamma }(1,1)\left\{ \begin{array}{ccc}
J_{c} & j_{2} & J_{b}\\
J_{c} & j_{2} & J_{b}\\
0 & k_{2} & k_{b}
\end{array}\right\} \\
 & \times \left\{ \begin{array}{ccc}
J_{b} & j^{i}_{1} & J_{a}\\
J_{b} & j^{j}_{1} & J_{a}\\
k_{b} & k_{1} & k_{a}
\end{array}\right\} C_{000}^{k_{b}k_{1}k_{a}}C_{000}^{0k_{2}k_{b}}\\
 & \times \left| \left\langle J_{c}\right\Vert j_{2}\left\Vert J_{b}\right\rangle \right| ^{2}\left| \left\langle J_{b}\right\Vert j^{i}_{1}\left\Vert J_{a}\right\rangle \right| \left| \left\langle J_{b}\right\Vert j^{j}_{1}\left\Vert J_{a}\right\rangle \right| c_{k_{2}0}(j_{2}j_{2})\\
 & \times \left[ (-1)^{j^{i}_{1}}c_{k_{1}0}(j_{1}^{i}j^{j}_{1})+(-1)^{j^{j}_{1}}c_{k_{1}0}(j_{1}^{j}j^{i}_{1})\right] P_{k}(\cos \theta )\, .
\end{array}
\end{equation}
 The coefficient \( k \) is even and satisfies the triangle rule \( \mathbf{k}=\mathbf{k}_{1}+\mathbf{k}_{2} \). 

We have calculated the theoretical values of the angular correlation function
for two possible cases. The electric field vector is along the \( x \) axis
and a perpendicular plane geometry is used to describe the process. We choose
the collision frame \( x,y,z \) attached to the sample and the \( z \) axis
coincides with the direction of incident the photon beam. Since the Stokes parameters
\( S_{1} \) and \( S_{2} \) refer to the same quantity, but with differently
oriented axes, one can take \( S_{2}=0 \). This is done by selecting the \( x \)
axis of the collision frame to coincide with the direction of maximum linear
polarization, i.e. the major axis of the polarization ellipse. The same polar
and azimuthal angles are chosen in the experiment. Here \( e_{1} \) and \( e_{2} \)
are the directions of propagation of the photo-electron and the Auger electron
respectively. (i) The photo-electron is observed in a fixed direction and the
Auger electron spectrometer is turned around to get the angular distribution
of the Auger electron with respect to the photo-electron. Here the maximum allowed
value of \( k \) is \( 2j_{2} \). This is our first case. (ii) The second
case is the complementary case i.e. the Auger electron is observed in a fixed
direction and the photo-electron spectrometer is turned around to get the angular
distribution of the photo-electron with respect to the Auger electron. Here
the maximum allowed value of \( k \) is \( 2j_{1,\max } \). This value gives
the highest order of the Legendre polynomials occurring in the correlation function.

The angular correlation function in DPI due to polarized photons now becomes
\\
i) Case 1 : the photo-electron is observed in a fixed direction and the Auger
electron spectrometer is turned around to get the angular distribution of the
Auger electron with respect to the photo-electron,

\begin{equation}
\label{11}
W(\theta )\sim 1+1.314P_{2}(\cos \theta )+1.100P_{4}(\cos \theta )\, .
\end{equation}
 ii) Case 2 : the Auger electron is observed in a fixed direction and the photo-electron
spectrometer is turned around to get the angular distribution of the photo-electron
with respect to the Auger electron,

\begin{equation}
\label{12}
W(\theta )\sim 1+0.817P_{2}(\cos \theta )+0.602P_{4}(\cos \theta )+0.570P_{6}(\cos \theta )\, .
\end{equation}
 If the incident photon is unpolarized Eq.(\ref{5}) reduces to Eq.(25) of reference
{[}\ref{we1}{]}. 

The solid line in the figures represents the theoretical curve and the dotted
points are experimental values of \( W(\theta ) \). Fig 1 gives the polar plot
of the angular correlation function for unpolarized photons. Figures 2 and 3 for
linearly polarized photons.

\vspace{0.3cm}
{\par\centering \includegraphics{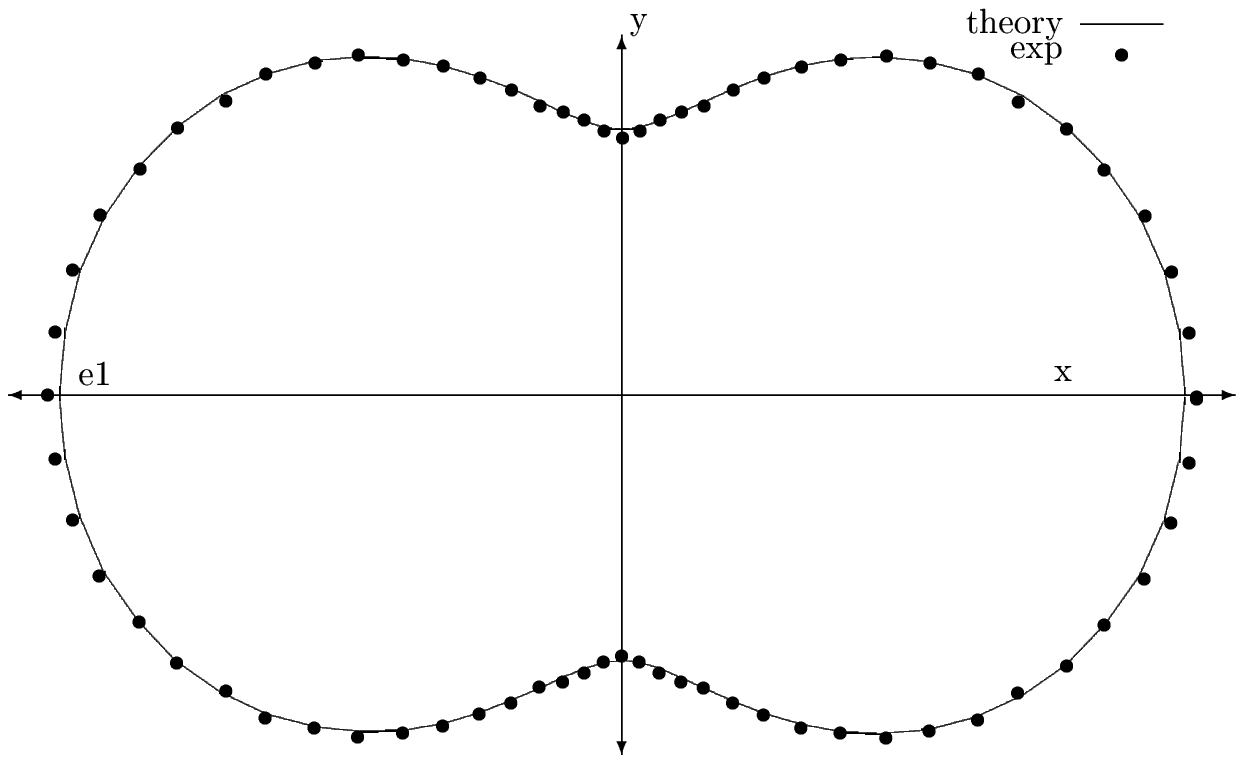} \par}
\vspace{0.3cm}

{\par\centering Fig.1: Polar plot of angular correlation function \( W(\theta ) \)for
2-step double photoionization of xenon due to unpolarized photon beam of \( 94.5\, eV \)
(\( 4d_{5/2} \) photoionization followed by \( N_{5}-O_{2,3}O_{2,3}\, ^{1}S_{0} \)
Auger decay). The photo-electron is observed in a fixed direction.\par}

\vspace{0.3cm}
{\par\centering \includegraphics{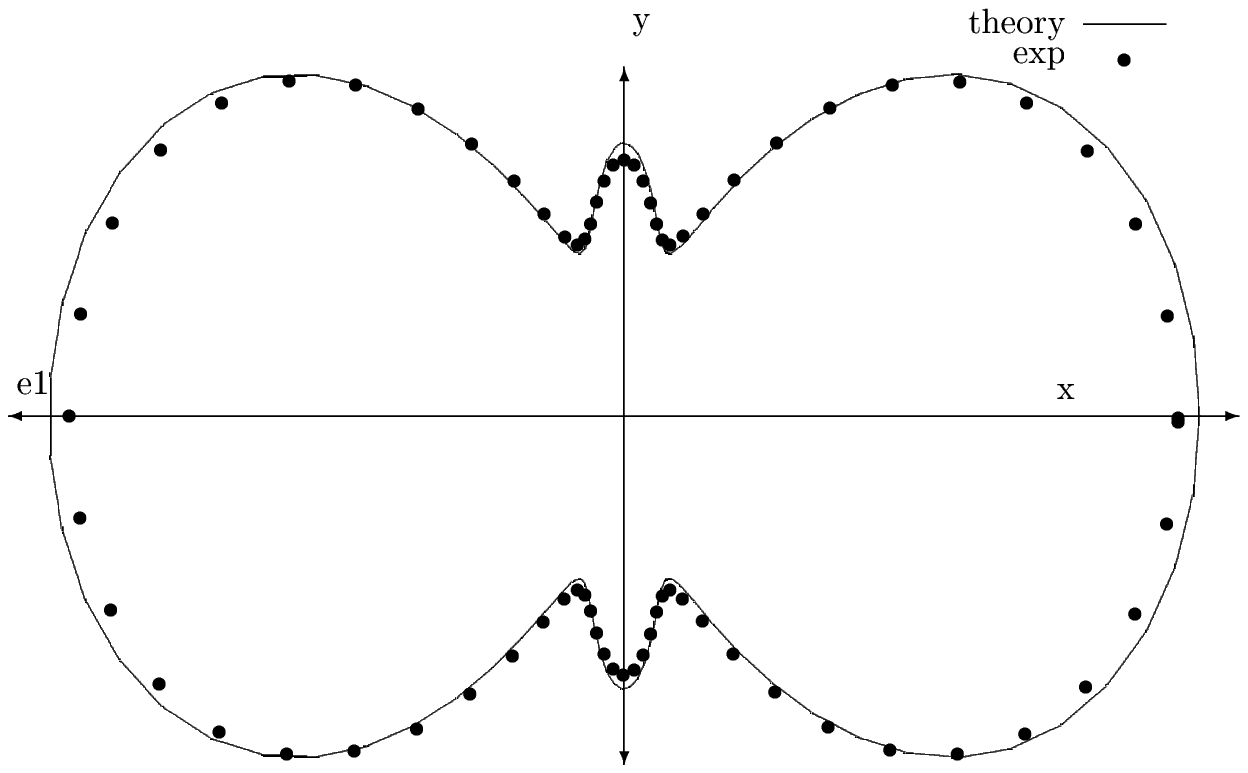} \par}
\vspace{0.3cm}

{\par\centering Fig.2: Angular correlation pattern for 2-step double photoionization
of xenon due to a linearly polarized photon beam of \( 94.5\, eV \) (\( 4d_{5/2} \)
photoionization followed by \( N_{5}-O_{2,3}O_{2,3}\, ^{1}S_{0} \) Auger decay).
The photo-electron is observed in a fixed direction.\par}

\vspace{0.3cm}
{\par\centering \includegraphics{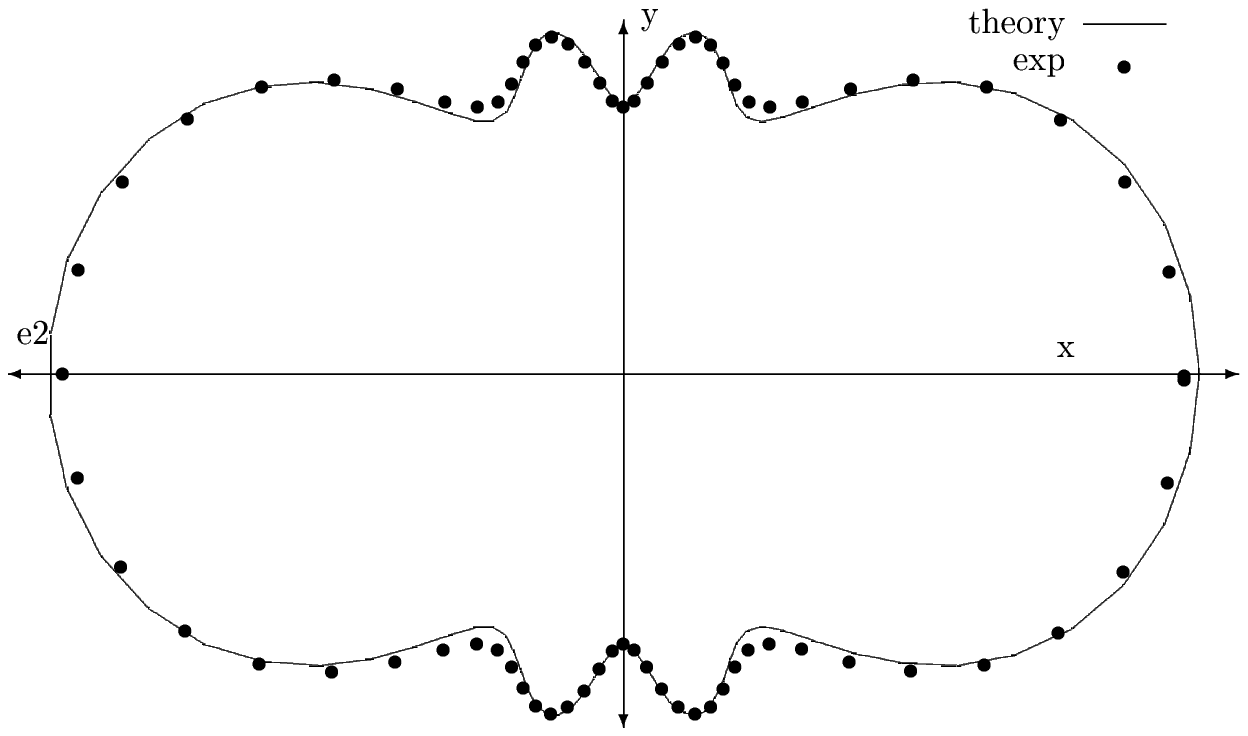} \par}
\vspace{0.3cm}

{\par\centering Fig.3 : Angular correlation pattern for 2-step double photoionization
of xenon due to a linearly polarized photon beam of \( 94.5\, eV \) (\( 4d_{5/2} \)
photoionization followed by \( N_{5}-O_{2,3}O_{2,3}\, ^{1}S_{0} \) Auger decay).
The Auger electron is observed in a fixed direction.\par}

\vspace{0.1in}
\textbf{Acknowledgement} : One of the author (CS) would like to acknowledge
the financial support provided by the University Grants Commission of India.

\end{document}